\documentclass[final]{IEEEtran}

\usepackage{amsthm,amssymb,graphicx,multirow,amsmath,color,amsfonts}
\usepackage[update,prepend]{epstopdf}
\usepackage[noadjust]{cite}
\usepackage[latin1]{inputenc}
\usepackage{tikz}
\usetikzlibrary{arrows,calc}		
\usepackage{bbm} 
\usepackage{pdfpages}
\usepackage{tabulary}
\usepackage{multirow}
\usepackage{comment}
\usepackage{subcaption}
\usepackage[font=small]{caption}
\usepackage{xcolor}
\usepackage{soul}

\def\nb0{{\mathbf{0}}}
\def\nb1{{\mathbf{1}}}

\allowdisplaybreaks 

\begin{document}
\graphicspath{{./Figures/}}
\title{
NTN-based 6G Localization: \\
Vision, Role of LEOs, and Open Problems  
}
\author{
Harish K. Dureppagari, Chiranjib Saha, Harpreet S. Dhillon, R. Michael Buehrer
\thanks{H. K. Dureppagari, H. S. Dhillon, and R. M. Buehrer are with Wireless@VT, Department of ECE, Virginia Tech, Blacksburg, VA 24061, USA. Email: \{harishkumard, hdhillon, rbuehrer\}@vt.edu. C. Saha is with the Qualcomm Standards and Industry Organization, Qualcomm Technologies Inc., San Diego, CA 92121, USA. Email: csaha@qti.qualcomm.com. The support of the US NSF (Grants 1923807 and CNS-2107276) is gratefully acknowledged.
}
\vspace{-1mm}
}

\maketitle

\begin{abstract}
Since the introduction of 5G Release 18, non-terrestrial networks (NTNs) based positioning has garnered significant interest due to its numerous applications, including emergency services, lawful intercept, and charging and tariff services. This release considers single low-earth-orbit (LEO) positioning explicitly for {\em location verification} purposes, which requires a fairly coarse location estimate. To understand the future trajectory of NTN-based localization in 6G, we first provide a comprehensive overview of the evolution of 3rd Generation Partnership Project (3GPP) localization techniques, with specific emphasis on the current activities in 5G related to NTN location verification. We then delineate the suitability of LEOs for location-based services and emphasize increased interest in LEO-based positioning. In order to provide support for more accurate positioning in 6G using LEOs, we identify two NTN positioning systems that are likely study items for 6G: (i) multi-LEO positioning, and (ii) augmenting single-LEO and multi-LEO setups with Global Navigation Satellite System (GNSS), especially when an insufficient number of GNSS satellites (such as 2) are visible. We evaluate the accuracy of both systems through a 3GPP-compliant simulation study using a Cram\'{e}r-Rao lower bound (CRLB) analysis. Our findings suggest that NTN technology has significant potential to provide accurate positioning of UEs in scenarios where GNSS signals may be weak or unavailable, but there are technical challenges in accommodating these solutions in 3GPP. We conclude with a  discussion on the research landscape and key open problems related to NTN-based positioning.
\end{abstract}


\section{Introduction} \label{sec:intro}
Satellite communication networks have been largely inaccessible for mass-market usage because of limited scale, cost, proprietary technology, and specialized equipment. However, the emergence of the 5G ecosystem has attracted the attention of the satellite community, leading to investigations in the 3rd Generation Partnership Project (3GPP) related to the use of satellites in public safety applications, fixed wireless access, and narrow-band internet-of-things (NB-IoT) connections. Despite this progress, significant challenges remain in order to utilize satellite communications as a core feature in 5G. 5G satellite communications, also termed non-terrestrial networks (NTNs) in 3GPP, have the potential of enabling truly global coverage, extending from emergency texting capabilities to delivering true mobile broadband. Positioning in 5G is crucial in many industries such as emergency services, logistics, and autonomous driving~\cite{3gpp22071}. While cellular positioning has evolved significantly, NTN-based positioning in Release 18 is still in the early stages and is limited to user equipment (UE) location verification~\cite{3gpp::location::verification::TR}. We believe that future releases of 5G and 6G will consider NTN positioning with low-earth-orbit (LEO), medium-earth-orbit (MEO), and geostationary-earth-orbit (GEO) satellites. This paper provides a vision of the extension of cellular positioning procedures, currently designed for terrestrial networks (TNs), to NTN, which could be considered in future releases of 5G and 6G. Note that the terms LEOs and LEO satellites are interchangeably used in the paper.

Not surprisingly, there have been some recent efforts in exploring the use of NTNs for positioning. In~\cite{PsiakiLeo2021}, the authors propose a method using a point-solution approach for LEO-based positioning that was originally developed for the Doppler-shift-based Global Navigation Satellite System (GNSS). However, this technique requires at least eight LEOs to be visible in order to perform a 3D location estimation. In~\cite{BENZERROUK2019496}, the authors propose a method for estimating the position and speed of a target using distributed Doppler information from the Iridium LEO constellation, employing nonlinear least squares and filtering algorithms for search and rescue services. In~\cite{HkUavLoc2023}, the authors proposed using UAVs as a means of realizing NTN-based accurate localization of emergency response personnel. In~\cite{ZakKassas2023}, the authors presented experimental results and receiver designs for navigation using multi-constellation LEO signals of opportunity. While these works provide several useful insights, they are largely disconnected from standardization activities. {\em We fill this gap by discussing current 3GPP efforts related to NTN positioning and sharing our vision on the evolution of this area as we prepare to enter the Beyond 5G and 6G era.}

\begin{figure*}
    \centering
    \includegraphics[width=.88\textwidth]{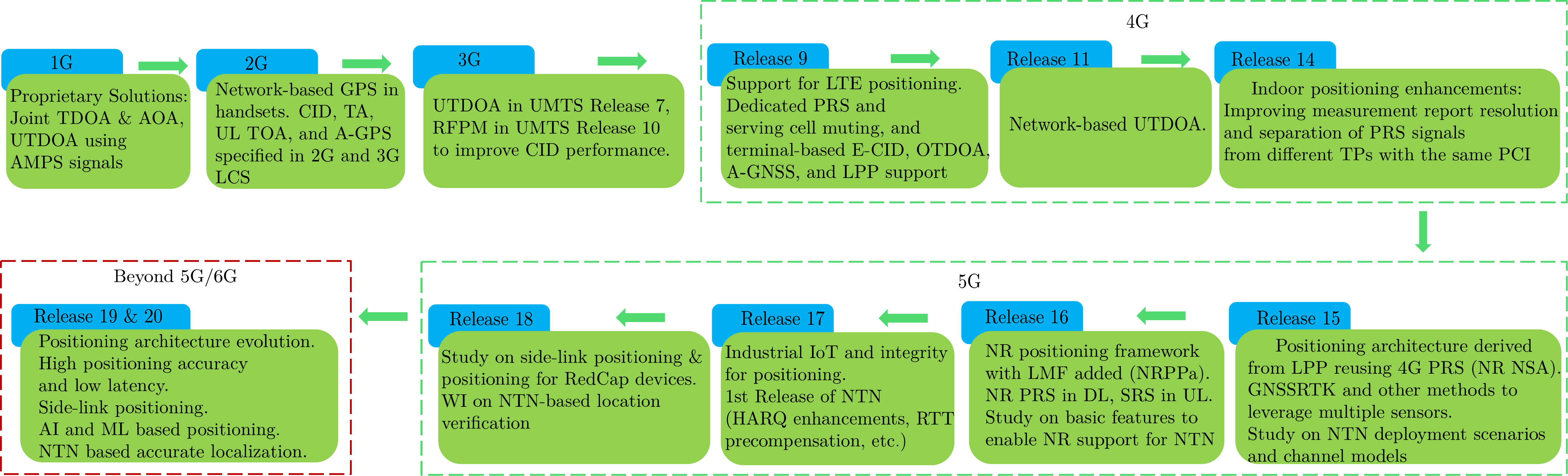}
    \caption{The evolution of positioning support pre and post-formation of 3GPP, see~\cite{del2017survey} for a detailed survey.}
    \label{fig:posevolution}
\end{figure*} 

In order to provide a comprehensive understanding of the current and future landscape of positioning systems in cellular networks and NTNs, we begin by providing a detailed overview of the evolution of positioning systems across different cellular standards, with specific emphasis on the current 5G activities on NTN-based location verification that requires a coarse location estimate. Next, we outline the evolution of traditional NTN navigation systems, the relevant support in 3GPP, and highlight the growing interest and suitability of LEOs for location-based services. To extend NTN support for accurate positioning, we identify and investigate two prospective positioning frameworks that are likely study items for 6G: multi-LEO positioning and augmenting single-LEO and multi-LEO setups with GNSS. To evaluate the performance of these systems in comparison to 5G location verification, we conduct an extensive simulation study using a Cram\'{e}r-Rao lower bound (CRLB) analysis. Additionally, we discuss some of the challenges specific to 3GPP standards that need to be addressed in order to enable these technologies in 6G. Finally, we highlight open problems that arise when LEOs are utilized for positioning, as they were originally deployed for communication purposes.

\section{Positioning in Terrestrial and Non-Terrestrial Networks}
\subsection{Cellular Positioning}
To put the contribution of this article into perspective, it will be useful to discuss the evolution of positioning methods in cellular networks (see Fig.~\ref{fig:posevolution}). In 1G, positioning was limited to intelligent vehicle highway systems and emergency services, utilizing joint time-difference-of-arrival (TDOA) and angle-of-arrival (AOA) for location estimation. During 2G, the Federal Communications Commission (FCC) in the United States (US) approved location requirements for 911 emergency calls, leading to the support of the Global Positioning System (GPS) in handsets. Additionally, assisted GPS was introduced in 2G to aid GPS with navigation messages. Before the emergence of 3G, European and American standardization groups collaborated to develop a functional description of location services (LCS) for 2G and 3G networks and positioning schemes such as cell-ID (CID), timing advance (TA), uplink time-of-arrival (UL TOA), and enhanced observed time difference (E-OTD) were specified. Later in 3G, uplink TDOA (UTDOA) and RF pattern matching (RFPM) were added to the supported methods to improve the CID positioning performance. Although 5G cellular positioning methods retain some fundamental elements from the past generations, from the architecture and protocol standpoint, 4G marked a significant shift in evolution. 4G led to the introduction of dedicated positioning reference signals (PRS) with serving cell muting, and different positioning methods like enhanced CID (E-CID), observed TDOA (OTDOA), UTDOA, and Long-Term Evolution (LTE) positioning protocol (LPP)~\cite{3gpp36355}.

 An LCS session may be initiated upon a location request from a UE or external client via the gateway mobile location center (GMLC), or core network (CN) node e.g. access and mobility management function (AMF)~\cite{3gpp23273}. The LMF interacts with the UE using LPP, and gNBs using New Radio Positioning Protocol Annex (NRPPa). The UE receives the required radio configuration from the next-generation radio access network (NG-RAN) over radio resource control (RRC) via the NR-Uu interface. Leveraging the centralized unit (CU) and distributed unit (DU) split introduced in 5G NR, both LPP and NRPPa protocols are transported over the control plane of the NG interface (NG-C) via AMF. The gNB CU is connected to a gNB DU over the F1 interface. The transmission/reception point (TRP) in 5G-RAN, a central part of a gNB DU, is an abstract node that can transmit PRS in downlink (DL) and perform sounding reference signals (SRS) based measurements in UL. Using the DL and UL measurements from the UE and TRPs, the LMF may compute the UE location using radio access technology (RAT) dependent positioning methods like TDOA, multi round-trip-time (RTT), DL AOA, UL AOA, and E-CID. 

In 5G Release 18, 3GPP continues NR positioning evolution by studying solutions for side-link positioning, extending positioning support for reduced capability UEs, and further improving positioning accuracy.  From Release 17, 3GPP has started standardizing enhanced mobile broadband (eMBB) and IoT services over NTN. Although NTN positioning is still in its early stages, it is worth noting that 3GPP has initiated discussions on using NTN for UE location verification~\cite{3gpp::location::verification::TR}. The objective of this work item is to verify the UE location up to 10km location accuracy without relying on UE-reported GNSS measurements, which we revisit in the next section.

{\subsection{Satellite-based Positioning}} 
We need to differentiate between cellular-based positioning methods and GNSS systems. The cellular positioning with its infrastructure can work as a standalone positioning unit without UE GNSS implementations. However, the primary applications requiring LCS  to date largely rely on UE GNSS. 
The evolution of GNSS systems began in the late 1980s with the US launching GPS. Different countries have their own GNSS systems, such as GLONASS by Russia, Galileo by the European Union, and BeiDou by China. GPS in particular operates in two primary frequencies called L1 (1575.42MHz) and L2 (1227.60MHz) catering to military and civilian applications. It uses direct sequence spread spectrum (DSSS) in which the information bits are modulated with a spreading code that spreads over the intended bandwidth ($\sim$1MHz for civilian access codes). To assist the UE with faster GNSS acquisition, support for GNSS in 3GPP started in Release 7 with A-GPS followed by assisted GNSS (A-GNSS) in Release 9. It was expanded in later releases with enhancements for improved positioning accuracy, support for various constellations, better indoor positioning, and improved security.

In this paper, we focus on \textit{extending cellular positioning to NTN}. As 3GPP primarily focuses on LEOs for NTN integration into cellular networks, our scope for NTN-based positioning further boils down to LEO-based positioning. We envision LEOs as the next frontier in cellular positioning for several reasons: (i) growing commercial interest and investments in deploying LEOs for communication purposes, (ii) availability of larger communication bandwidth resulting in faster acquisition time and improved accuracy, (iii) no additional UE implementation complexity, and (iv) LEOs orbit at much lower altitudes compared to GNSS, offering significant advantages such as better link budget and evolving satellite geometry due to dynamic satellite movement. However, there are also challenges due to the high mobility of LEOs such as high Doppler and frequent handovers between different satellite groups to maintain continuous coverage. Despite these challenges, LEO-based positioning can be a potential alternative to GNSS, especially in challenging scenarios such as dense urban or indoor environments where GNSS signals are too weak due to non-line-of-sight (NLOS) propagation.\\

\begin{figure*}[t!]
\centerline{\includegraphics[width=.87\textwidth]{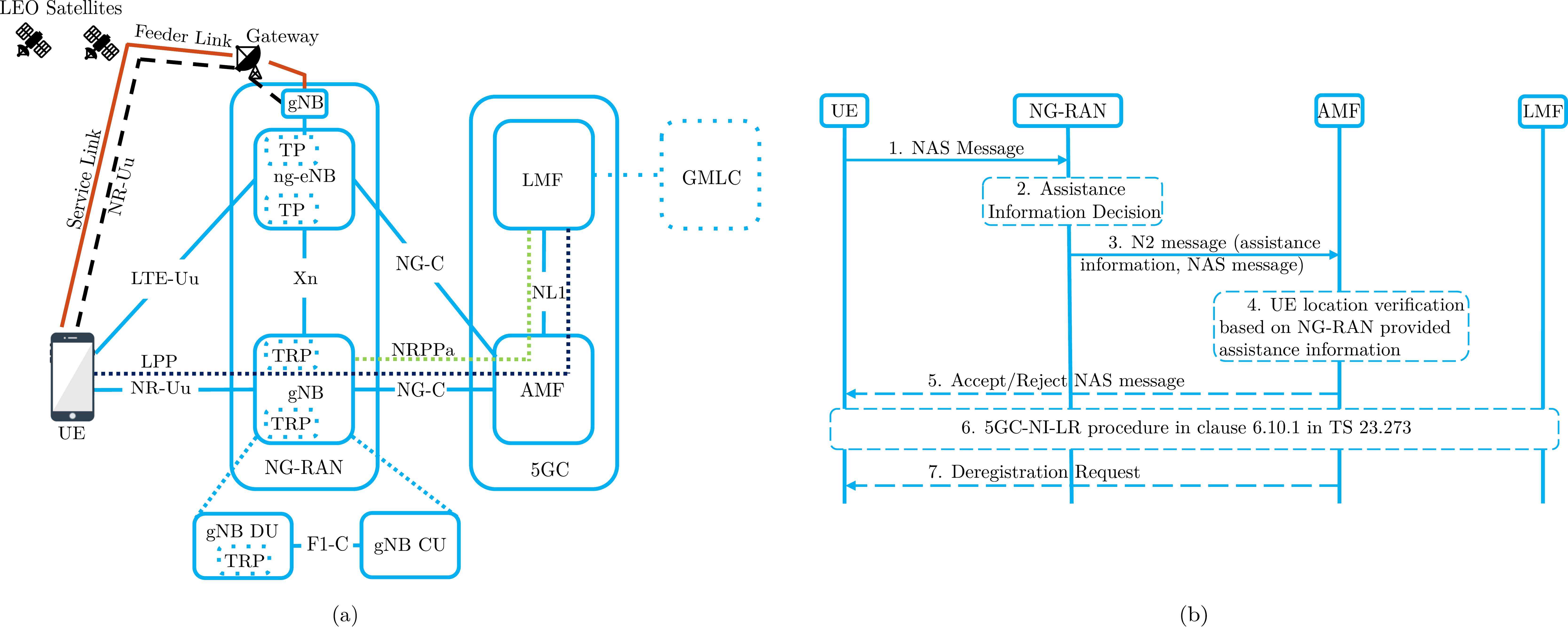}}
\caption{(a) 5G positioning architecture with our envisioned future extensions to NTN. (b) A location verification solution discussed in 3GPP system architecture (SA) working group (WG) using the 5G positioning architecture.}
\label{fig:5gposarchitecture}
\end{figure*}

\section{From 5G to 6G NTN:\\ New Frontiers for Positioning}
\subsection{NTN Framework in 5G}
Starting from 3GPP Release 15, the integration of satellite networks into the 5G ecosystem was initiated through the engagement of the satellite industry in the 3GPP standardization process. NTN is becoming an integral part of the 5G ecosystem due to its potential of extending the reach of 5G TNs to under-served areas, thereby improving the reliability and resiliency of the network, particularly against physical attacks and natural disasters. Other key uses of NTN in 5G include offering 5G services to users onboard airborne vehicles, enabling wide-area IoT services, and relaying emergency and public safety messages. NTN refers to a range of satellite and airborne vehicle networks, including LEOs, MEOs, GEOs, High Altitude Platform Systems, and air-to-ground networks. LEOs are particularly attractive due to their real-world significance and the possibility of deploying mega-constellations with reusable technologies. 

In NTN access, a satellite may implement either a bent pipe payload (on-ground NTN gNB) or a regenerative payload (onboard NTN gNB). The positioning architecture in Fig.~\ref{fig:5gposarchitecture} assumes a bent pipe payload. Notably, the essence of positioning lies in the transmit-receive time difference measurements taken at TRP, which could be an on-ground/onboard NTN gNB or any reference point designated by the network. Moreover, the network is capable of closely tracking the satellites, hence the feeder link delay. Thus, the choice of payload type does not significantly impact the overall accuracy of NTN-based positioning techniques.
In releases 15 and 16, the RAN identified several technical challenges related to the deployment of NTN, such as higher propagation delays in NTN compared to TNs. This led to new designs maintaining large TAs, enhancing random access (RA) procedures,  and hybrid automatic repeat request (HARQ) processes. See~\cite{NTN5g26g} for more details on the radio aspects of satellite access. 

{\em An important aspect of NTN design is reliance on the GNSS capabilities of a UE.} To close a radio link with a satellite, a UE must calculate the relative speed and RTT of the service link, for timing and Doppler resolution. This can be done by acquiring the satellite ephemeris at the gNB and broadcasting it as system information, and the estimate of UE location. 
Evidently, unlike TN, UE location resolution is a central component of NTN access. Hence,  GNSS capability is mandatory for NTN access, and UEs without GNSS are out of the scope of 5G NTN. As we will discuss further, this sets the stage for a new set of design problems where UEs without GNSS can be localized using standalone RAT-dependent methods and hence can connect to NTN.

\subsection{NTN-Based Location Verification}
Although NTN positioning is not in the direct scope of 5G, the requirement for designing RAT-dependent positioning solutions for NTN is motivated by the requirement of UE location verification. Location verification is vital in cellular networks for various reasons, such as public land mobile network (PLMN) selection, emergency calls, subscription and billing, and lawful intercepts. For most such purposes, it is enough to use CID-based positioning since a cellular cell may cover only a few kilometers (e.g. 2km) and this accuracy may be enough for these purposes. For NTNs, however, each cell may cover hundreds of kilometers, rendering this approach inappropriate. We illustrate some example scenarios as follows.

{\em CN Selection.} 
After a UE connects with the RAN (i.e. moves to the {\em RRC\_Connected} state), the network selects the appropriate CN for the UE, taking into account various factors including the selected PLMN ID and location information such as the serving cell of the UE~\cite{3gpp38300}. For NTN, the cells can be moving and significantly larger than those of TNs, spanning vast areas that may even cover multiple countries with different CNs connected to the same NTN RAN. This can make it challenging to determine the appropriate CN for a connected UE, particularly when the UE is located close to a country border and the serving cell does not provide sufficient granularity. Additionally, malicious UEs may attempt to fake their selected PLMN and GNSS measurements to connect to a different CN. Hence, it is crucial for TNs to have the necessary infrastructure to verify UE location without relying solely on UE-reported information. The same motivation drove 3GPP to introduce NTN-based location verification in 5G.

{\em Regulatory Aspects.}
The network operators must reliably know the location information of a UE to provide services that comply with the governing national or regional regulatory requirements such as public warning systems, charging and billing, emergency calls, lawful intercept, and data retention policy in cross-border scenarios and international regions. 

As UE-reported location can be easily tampered with, 3GPP is working on designing a RAT-dependent network-based UE location verification system that provides a horizontal accuracy of 10km~\cite{3gpp::location::verification::TR}. The network verification procedure is illustrated in Fig.~\ref{fig:5gposarchitecture}(b). The AMF, based on the assistance information provided by the satellite (NG-RAN) decides whether to trigger a location verification process (step 4). The AMF might take different actions based on the status of UE location verification, for instance, rejected CN access or denied service requests for calls or data (step 5). The LMF initiates UE positioning through network-induced location request (NI-LR) (step 6). 
During this step, the LMF does not rely on the location information provided by the UE. Based on the outcome of step 6, the AMF may decide to deregister the UE from CN (step 7). In step 6, the network may leverage any RAT-dependent positioning method. Since the RAT has moving TRPs (i.e. satellites) in this case, the positioning methods could be significantly different from TNs, which we discuss next.

\subsection{Perspective of Positioning Methods in NTN} \label{sec:ITbounds}
In this section, we discuss how positioning methods in TNs may differ when integrating NTN into cellular networks. {\em Although satellite-based positioning has been studied extensively in the past, it is to be noted that fitting these solutions in the 5G NTN framework which is primarily designed for communication is not straightforward.} In the literature, various positioning methods based on delay and Doppler measurements, e.g.  TOA, TDOA, RTT or two-way TOA, and frequency-difference-of-arrival (FDOA) have been discussed. The choice of positioning method depends on the specific requirements of the system and the available resources. Note that 5G positioning has the capability and signaling support to accommodate most of the delay-based positioning methods but not any Doppler-based positioning methods such as FDOA.

When it comes to NTN-based positioning, there are several factors to consider. For instance, in TDOA, the geometric dilution of precision (GDOP) plays a critical role in minimizing position errors. However, given that early NTN deployments are primarily coverage-centric, only single LEO-based positioning techniques are prioritized in 5G Release 18~\cite{XlinRel18}. Multi-satellite techniques require denser constellations and satellites with overlapping coverage, which will be in the second phase of NTN to enhance capacity. Therefore, to perform TDOA with a single LEO, the LEO must move significantly in its trajectory between TDOA measurements, which requires high measurement time. During this time, it can be challenging to maintain a stable clock (i.e. a clock with constant offset) at the UE. Although FDOA has been considered for LEO-based positioning, e.g., see~\cite{PsiakiLeo2021}, its implementation poses several challenges. First, it requires at least 8 visible LEOs to perform accurate 3D estimation which is not always guaranteed. Second, the need for a significant redesign of the existing PRS to provide sufficient Doppler resolution for accurate FDOA measurements. To address these concerns, RTT-based positioning for NTN is being discussed in the 3GPP RAN WGs~\cite{3gpp::location::verification::TR}. This method involves DL TOA measurements using PRS transmitted from LEO and UL TOA measurements using SRS transmitted from the UE. By taking several two-way TOA measurements as the LEO moves in its orbit, the position of the UE can be calculated.

\begin{figure*}[htbp]
\centering
  \includegraphics[width=.84\textwidth]{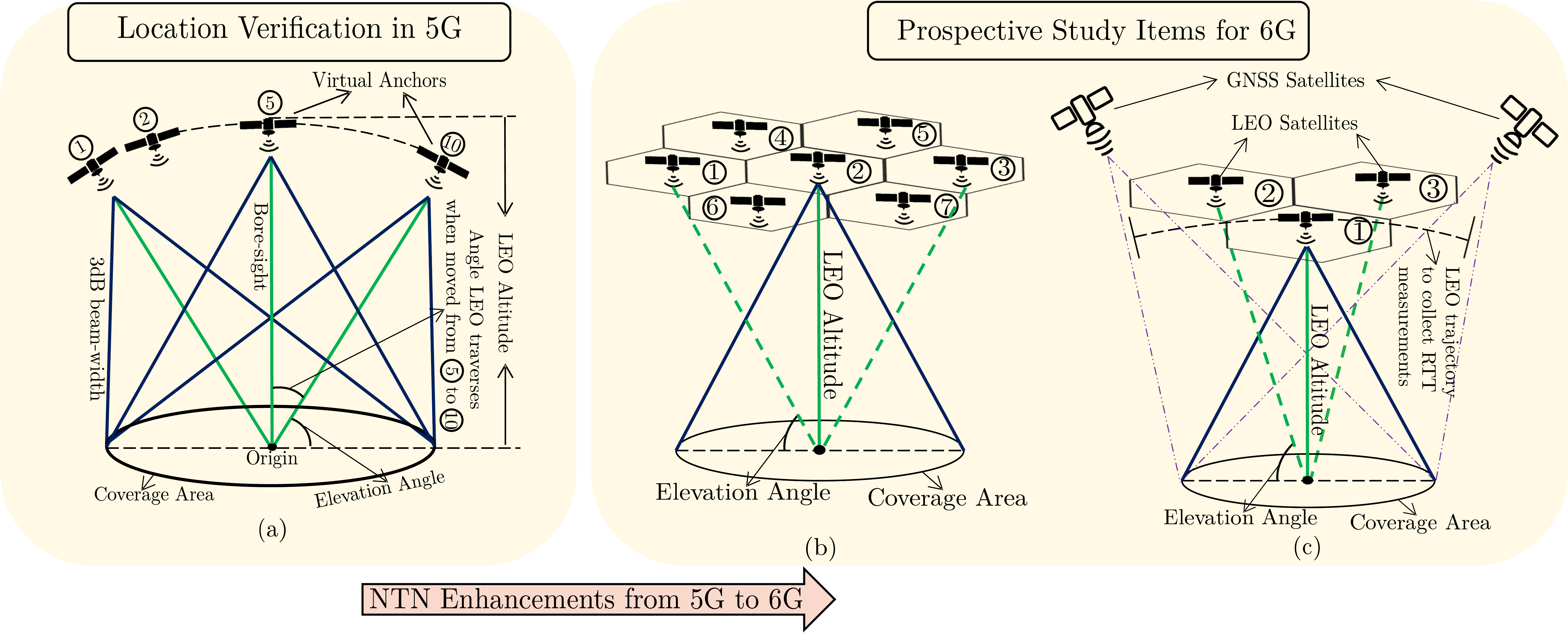}
        \caption{System model figures illustrating potential NTN enhancements from 5G to 6G: a) {\em Single LEO Positioning:} Positioning with a single LEO satellite with an earth-fixed beam. The coverage area corresponds to the 3dB beam-width of the satellite antenna pattern. As LEO moves in its orbit, 10 virtual anchors are specified where independent RTT measurements are collected. b) {\em Multi-LEO Positioning:} Positioning with LEOs. Seven LEOs form a hexagonal grid. UEs in the central beam of LEO 2 are considered for evaluation. TDOA measurements across LEOs and RTT measurements from the reference LEO. c) {\em GNSS + LEO(s) Positioning}: Positioning with 2 GNSS + one or more LEOs. Two GNSS satellites provide one TDOA measurement while LEOs provide TDOA and RTT measurements and a single LEO provides RTT measurements over the measurement time. The coverage area under consideration corresponds to the 3dB beam-width of the LEO satellite.}
        \label{fig:comb_sysmodel}
\end{figure*}

While RTT-based positioning serves as a baseline for single LEO positioning in 5G, multi-RTT for multiple LEOs poses certain challenges. Using two-way TOA necessitates compensating for propagation delays across several LEOs, which differ significantly. These differences may exceed tolerable delay spreads in 5G, making it insufficient to compensate for propagation delay only for the serving satellite, as in TNs. Further, GNSS is needed to enable both single LEO positioning and multi-RTT with multiple LEOs. While the value of LEO positioning might be questioned with UE equipped with GNSS support, it is crucial to understand that LEO positioning in 5G focuses on location verification, not explicit positioning.
Besides, as we will discuss in Section~\ref{sec:research-landscape}, it is natural to wonder whether a standalone cellular positioning solution using LEOs can serve as an alternative to GNSS or enhancement of GNSS, especially indoors.

\section{NTN-based Positioning Case Studies for 6G} \label{sec:SysMod_NumRes}
In this section, we evaluate the NTN positioning performance achieved by LEOs. In addition to the NTN-based location verification use case currently being studied in 3GPP RAN WG, we study two important use cases that have been identified in this article as two likely work items beyond 5G and 6G, namely, multi-LEO positioning, and augmenting single-LEO and multi-LEO setups with GNSS, which is especially relevant when an insufficient number of GNSS satellites are visible to provide an unambiguous position fix. 
 From the system design perspective, our simulation results pave the way for a full-blown NTN-based positioning system to be built into cellular systems. For the evaluation of positioning performance, we use CRLB  which provides a position error bound (PEB) that can be obtained using these methods. We consider both TDOA and RTT-based positioning methods under realistic propagation assumptions. The large-scale parameters of the radio channel between UE and LEO were derived from the agreed channel model for 3GPP evaluations~\cite{3gpp38811}, and LEO link budget numbers and multi-LEO constellation were derived from~\cite{3gpp38821}. Further, it is worth noting that within the considered measurement time window, and with the careful selection of LEOs positioned at relatively high elevation angles, the geometry of LEOs remains relatively stable and less susceptible to abrupt changes.
 
\subsection{Single LEO Positioning}  \label{sec:singleleopos}
 Fig.~\ref{fig:comb_sysmodel}(a) presents the system model for single LEO-based positioning, where a LEO is positioned at a 600km orbit. The UEs are distributed within the 3dB beam-width of the LEO antenna pattern which is referred to as the coverage area. For simulation, we only consider the center beam of the satellite, i.e., when the beam center is at (or near) the nadir of the satellite. The line-of-sight (LOS) probability, shadowing,  and path loss for each link are determined by link distance and LEO elevation angle. A UE located at the center of the coverage area corresponds to an elevation angle of $90^\circ$ and a bore-sight angle of the beam pattern which is $0^\circ$. We define measurement time as the length of the time window to collect RTT measurements for a single UE at different LEO positions which serve as virtual anchors. The mean PEB is obtained by averaging PEB across all the UEs dropped inside the satellite coverage. For simulation, we considered the orbital velocity of $\sim7.6$km/s and the S-band (2GHz carrier frequency) with 10MHz DL and UL bandwidth. Furthermore, we have accounted for the PRS processing gain due to its wideband allocation. With a 10MHz bandwidth and assuming 15KHz sub-carrier spacing, the PRS signal can be sent on a maximum of 600 subcarriers, resulting in approximately 28dB gain.

\begin{figure}[t]
\centerline{\includegraphics[width=.82\columnwidth]{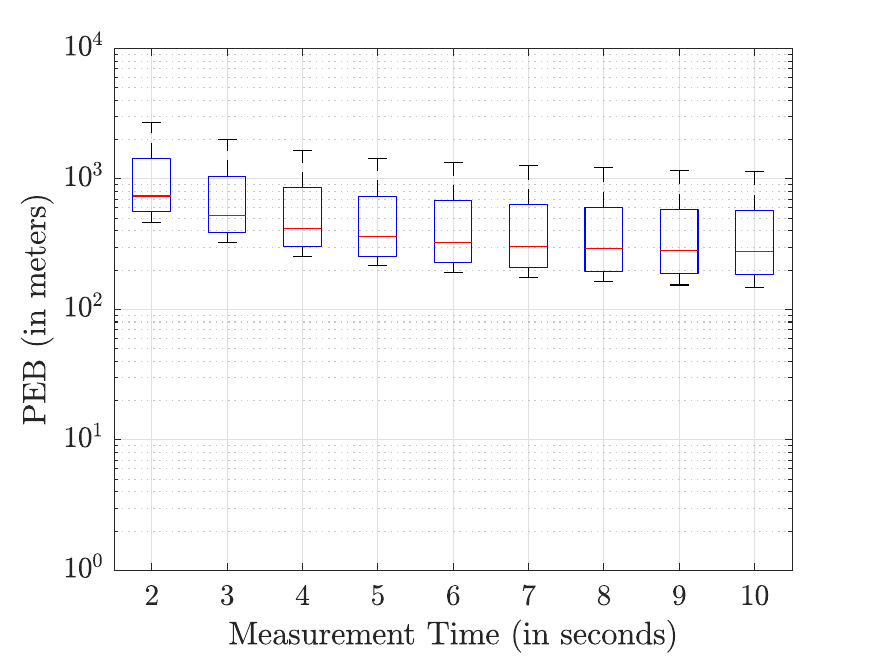}}
\caption{Box plot presents the PEB values obtained using single LEO (altitude of 600km) RTT measurements for various measurement times. The measurement times, in seconds, are [2 3 4 5 6 7 8 9 10], and the corresponding mean PEB values in meters are [2104.84, 1569.59, 1307.17, 1158.59, 1068.03, 1010.36, 973.69, 950.85, 934.97].}
\label{fig:singleleoboxplot}
\end{figure}

Fig.~\ref{fig:singleleoboxplot} presents the box plot of the PEB or position error values, obtained using single LEO RTT measurements for various measurement times. The trend in PEB suggests that longer measurement times result in more accurate position estimates because of more LEO position diversity. This result illustrates that a single satellite RTT is enough to satisfy the requirement of 10km positioning accuracy for network location verification. The significant difference between the mean and median PEB values indicates that for certain users, located near the satellite orbit and observe worse GDOP, the positioning accuracy is far worse than the central tendency. Although it is not clear how to handle such users if single LEO positioning is used in step 6 (Fig.~\ref{fig:5gposarchitecture}(b)), this makes a strong case for multi-LEO positioning, which we discuss next.

\subsection{Multi-LEO Positioning}  \label{sec:multileopos}
Fig.~\ref{fig:comb_sysmodel}(b) depicts a system model setup consisting of multiple LEOs.  For the simplicity of simulation, all LEO satellites are positioned at the same altitude ($780$km, similar to Iridium). The coverage area corresponds to the 3dB beam-width of the center satellite. We assumed a longitude gap of $13^\circ$ and a latitude gap of $6.9^\circ$ between any two satellites which results in the inter-satellite distance of $\sim$1633.4km.  The number of active satellites is assumed to be 3, 4, and 7. The satellite selection was performed to maximize GDOP. The center satellite is assumed to be the reference or {\em serving} satellite, while others are neighboring satellites.

\begin{figure}[t]
\centerline{\includegraphics[width=.82\columnwidth]{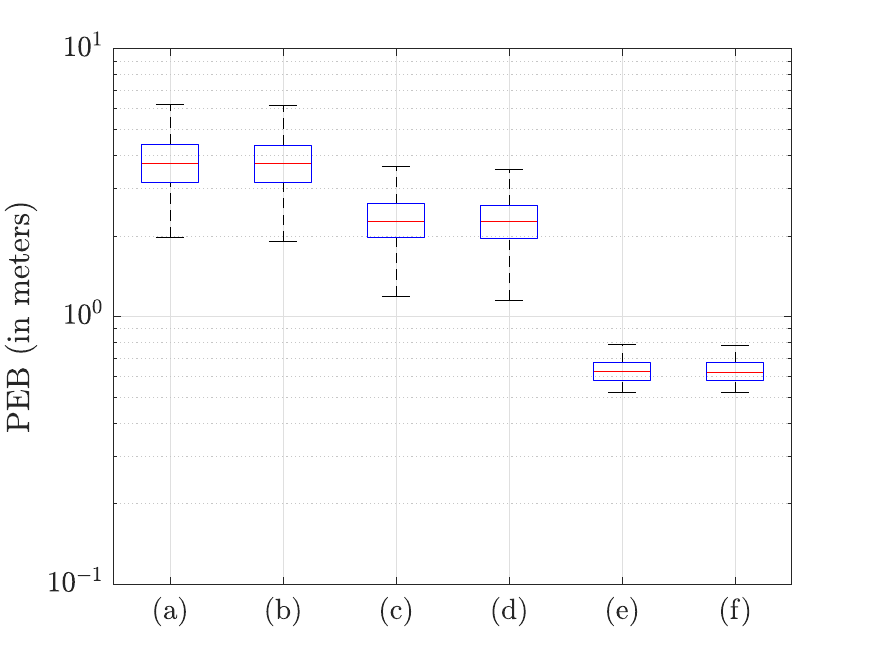}}
\caption{Box plot presents PEB values for 6 distinct scenarios involving multiple LEOs (altitude of 780km): (a) 3 LEOs with 2 TDOA measurements, (b) 3 LEOs with 2 TDOA + RTT measurements from the reference LEO, (c) 4 LEOs with 3 TDOA measurements, (d) 4 LEOs with 3 TDOA + RTT measurements from the reference LEO, (e) 7 LEOs with 6 TDOA measurements, and (f) 7 LEOs with 6 TDOA + RTT measurements from the reference LEO. The corresponding mean PEB in meters for these scenarios are [3.87, 3.84, 2.33, 2.32, 0.63, 0.63].}
\label{fig:multileoboxplot}
\end{figure}

\begin{figure}[t]
\centerline{\includegraphics[width=.82\columnwidth]{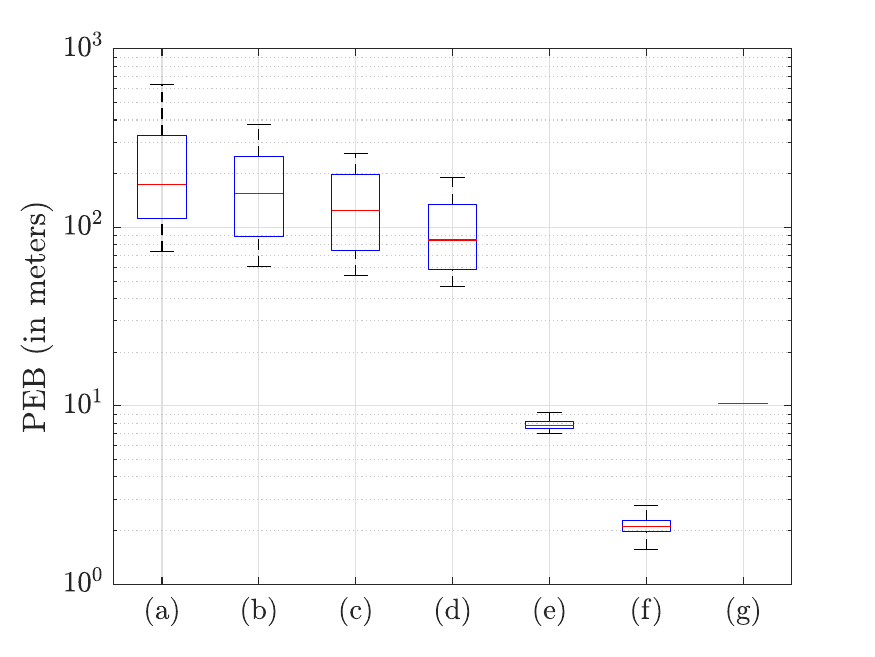}}
\caption{Box plot presents PEB performance for 4 different scenarios: (a)-(d) a single LEO (altitude of 780km) as an addition to 2 GNSS (altitude of 20200km) for different measurement times [2 5 7 10] seconds, (e) 2 LEOs as an addition to 2 GNSS, (f) 3 LEOs as an addition to 2 GNSS, and (g) 3 GNSS. TDOA measurements are obtained from GNSS and LEOs provide TDOA and RTT measurements. The corresponding mean PEB in meters for all scenarios are [243.10 171.37 135.75 97.71 7.90 2.14 10.35].}
\label{fig:gnss_leoboxplot}
\end{figure}

To evaluate multi-LEO positioning, we consider (i) TDOA across LEOs and (ii) RTT measurements from the reference LEO. It is worth noting that multi-LEO positioning is instantaneous whereas for a single LEO, RTT measurements are collected over measurement time. As demonstrated in Fig.~\ref{fig:multileoboxplot}, utilizing multiple LEOs can significantly improve the positioning accuracy over a single LEO. Specifically, with the sixth scenario, we are able to achieve sub-meter-level accuracy. This value is three orders of magnitude lower than the PEB with a single LEO, even with a measurement time of 10 seconds. Moreover, the number of outliers is significantly reduced compared to the single LEO. Our simulation clearly indicates that multiple LEOs can not only resolve the location verification problem currently discussed in 3GPP but can also be a promising solution for standalone NTN positioning.

\subsection{GNSS + LEO Positioning}  \label{sec:gnss_leo_pos}
In this section, we study localization using both GNSS and LEOs which is relevant when only a few GNSS satellites (such as two) are visible as shown in Fig.\ref{fig:comb_sysmodel}(c). We consider (i) TDOA across GNSS, (ii) TDOA across LEOs, and (iii) RTT from the reference LEO. For a single LEO, we collect RTT measurements over a certain measurement time.  For LEOs, the simulation assumptions are the same as those given in Section~\ref{sec:multileopos}. For GNSS, we considered an altitude of 20200km, an operating frequency of 1575.42MHz, GPS L1 coarse acquisition (C/A) or civilian code transmitted at 1.023Mbps, and link budget numbers as given in~\cite{Edkgps2005}. As illustrated in Fig.~\ref{fig:gnss_leoboxplot}, combining RTT measurements from a single LEO with GNSS measurements results in sub-100 m positioning accuracy. Positioning accuracy can be significantly improved by combining more than one LEO with the GNSS setup. Notably, for three GNSS, the variance of PEB is close to zero, because the flat wide beam of GNSS leads to uniform received signal-to-noise-ratio (SNR) across UEs dropped in a coverage area under consideration. Furthermore, we assume a 43dB processing gain for GPS L1 C/A code, which can be achieved after perfect synchronization. Note that, as LEOs can be facilitated with the NR NTN framework, the synchronization procedure and broadcasting assistance information can be done much faster than a typical GNSS cold start.

\section{Research Landscape and Open Problems} \label{sec:research-landscape}
 Many of the open research problems in LEO-based positioning stem from the fundamental fact that, unlike GNSS networks, the emerging mega-constellation LEO networks such as Starlink, Kuiper, and OneWeb, are designed for {\it communications} and not {\em geolocation}.  To this end, the satellites have multiple beams, and the adjacent beams are typically assigned different frequencies to reduce inter-beam (or inter-cell) interference. To maximize area spectral efficiency, the antenna beam-widths are kept relatively small. This design contrasts with optimum NTN deployment for positioning, where broad beam coverage and overlapping satellite beams are beneficial. That being said, there are certain aspects that need to be studied to maximize the localization capabilities of 6G NTN. 

 {\em Spare Beams.}
 Since PRS is designed to be decoded at very low SNR, the satellites may use {\em spare beams} to transmit PRS to locations outside its beam coverage. These spare beams can be activated opportunistically when the satellite has less ground area to cover (e.g., the LEO is passing over a land-ocean boundary). 
 
{\em New Reference Signal Design.}
  Another shortcoming with the current positioning process is the need for the UE to search for a wideband PRS within a specific time search window. In TNs, where distances between UE and TRPs are smaller, this window is shorter than a subframe duration. However, in NTN setups with larger relative distances between reference and neighboring TRPs, the search window expands across multiple subframes. For UEs, extended searches for wideband signals can be power-intensive. Designing quasi-co-located narrow-band synchronization signals, akin to synchronization signal blocks, alongside PRS in a deterministic manner might be advantageous. To enable Doppler measurements, 3GPP needs to standardize new narrowband PRS, with resources defined over multiple slots. The TRP and the UE need to maintain phase coherence over the transmission of this signal. While 3GPP RAN focuses on single satellite RTT, extending it to multi-RTT encounters fundamental challenges. For instance, accommodating multiple satellites necessitates UE adjustment of timing and frequency compensation for SRS reception. The specification needs to permit the UE to momentarily modify UL timing and frequency for SRS transmission while maintaining closed-loop timing and frequency compensation with the serving satellite.

{\em More Dynamic Signaling.}
  Since satellite locations are constantly changing, the nature of positioning assistance information sent by the LMF to UE should also change over time. For instance, assistance information related to the configuration of neighboring satellites signaled at one time may become invalid in the future. Similarly, the static PRS muting pattern in TN needs adjustment. In order to receive the PRS from a distant satellite, the transmission of PRS from a nearby satellite needs to be muted. However, since the satellites are moving, the muting patterns must also be dynamic in nature. In such scenarios, assuming overlapping PRS instances of multiple satellites, the serving satellite may need to mute adjacent symbols/slots of PRS instances to account for the larger propagation delay of distant satellite PRS.

  {\em LEO and GNSS.}
The above discussion assumes the use of LEO networks in isolation for positioning and navigation purposes. Another research avenue opened by NTNs is enhancing navigation through orbital diversity. For instance, our initial simulations show that LEOs can augment or enhance existing GNSS systems. This enhancement is particularly valuable in scenarios like dense urban areas, where GNSS signals often fail. From the cellular system design perspective, for UE-assisted positioning, the LMF needs to jointly process the GNSS and LEO measurements. Alternately, for UE-based positioning, the cellular modem needs to accumulate GNSS measurements and 5G NR radio measurements from LEO and process them jointly.  

In conclusion, NTN positioning is a practical and relevant research topic with many fundamental open problems. We hope that this article will put a spotlight on some of these as we start entering the beyond 5G and 6G era.

\bibliographystyle{IEEEtran}
\bibliography{ref}

\end{document}